\documentclass[11pt,aps,prd,preprint,superscriptaddress]{revtex4-1}
\usepackage{amsmath}
\usepackage[dvips]{graphicx}
\usepackage[english]{babel}
\begin{document}
\title{A thermodynamic motivation for dark energy}
\author{Ninfa Radicella\footnote{E-mail: ninfa.radicella@uab.cat} and
Diego Pav\'{o}n\footnote{E-mail: diego.pavon@uab.es}}
\affiliation{Departament de F\'{\i}sica, Universitat Aut\`{o}noma
de Barcelona, 08193 Bellaterra (Barcelona), Spain.}
\begin{abstract}
It is argued that the discovery of cosmic acceleration could have
been anticipated on thermodynamic grounds, namely, the generalized
second law and the approach to equilibrium at large scale factor.
Therefore, the existence of dark energy -or equivalently, some
modified gravity theory- should have been expected. In general,
cosmological models that satisfy the above criteria show
compatibility with observational data.
\end{abstract}
\maketitle
\section{Introduction}
The standard cold dark matter (SCDM) model \cite{peebles} was in
good health until around the last decade of the previous century
when it became apparent that the fractional density of matter
falls well below the Einstein-de Sitter value, $\Omega_{m} = 1$
-see e.g. \cite{mnrs-maddox,nature-efstathiou}. The death blow
came at the close of the century with the discovery of the current
cosmic acceleration \cite{riess}, something the said model cannot
accommodate by any means. However, to account for the acceleration
in homogeneous and isotropic models one must either introduce some
exotic energy component with a huge negative pressure (dubbed dark
energy) or, more drastically, devise some theory of gravity more
general than Einstein relativity \cite{reviews}. Thus, both
solutions appear somewhat forced and not very aesthetical. Here we
argue that dark energy (or something equivalent) is demanded on
thermodynamic grounds. In other words, we provide what we believe
is a sound thermodynamic motivation for the existence of dark
energy.

Our argument is based on that the natural tendency of systems to
evolve toward thermodynamical equilibrium is characterized by two
properties of its entropy function, $S(x)$, namely, it never
decreases, $dS(x)/dx \geq 0$, and is convex, $d^{2}S(x)/dx^{2} <
0$ \cite{callen}. In the context of an ever expanding
Friedmann-Roberson-Walker (FRW) cosmology this translates in that
the entropy of the apparent horizon plus that of matter and fields
enclosed by it must fulfill $S'(a) \geq 0$ at any $a$ -the
generalized second law (GSL)- as well as $S''(a) \leq 0$ as $a
\rightarrow \infty$, where $a$ is the scale factor of the FRW
metric and the prime means $d/da$. The apparent horizon in FRW
universes always exists (which is not generally true for the
particle horizon and the future event horizon) and is known to
posses not only an entropy proportional to its area
\cite{bak-rey,cai-2008} but also a temperature \cite{cai-2009}.
Besides, it appears to be the appropriate thermodynamic boundary
\cite{wang-2006}.

Before we proceed, it is fair to recall the (at least theoretical)
existence of systems lacking any global maximum entropy state,
such as Antonov's sphere \cite{lyndenbell68,lyndenbell99}. The
latter system consists in a sphere enclosing a number of particles
that share some total energy. If the sphere radius happens to
increase beyond some critical value, the system becomes unstable
and the entropy function ceases to have a global maximum. We shall
apply our argument under the assumption that the Universe tends to
a state of maximum entropy irrespective of whether it may
eventually reach it or not. It would be odd and frustrating that
such a universal principle as the second law of thermodynamics
could not be applied to the Universe as a whole, especially given
the close connection between thermodynamics and gravity
\cite{Ted,Pad}.

Section II illustrates why dark energy (or some or other modified
gravity model) is required on thermodynamic basis and study some
dark energy models to see whether they fulfill the thermodynamic
criteria. Section III applies the said criteria to some
representative modified gravity models. Finally, section IV
summarizes and discusses our findings. As is customary, a naught
subscript stands for the present value of the corresponding
quantity.

\section{Why dark energy was to be expected}
Let us consider a FRW universe. Its entropy is contributed by two
terms: the entropy of the apparent horizon which is proportional
to its area, ${\cal A} = 4 \pi \, \tilde{r}_{A}^{2}$, and the
entropy of the fluids enclosed by the horizon. Here $\tilde{r}_{A}
= (\sqrt{H^{2} \, + \, (k/a^{2})})^{-1}$ denotes the radius of the
horizon and $H$ the Hubble factor of the FRW metric
\cite{bak-rey}. As is well-known, $S_{A} \equiv \frac{k_{B}}{4}\,
\frac{\cal A }{\ell _{Pl}^2} $ where $\ell_{Pl}$ and $k_{B}$ stand
for Planck's length and Boltzmann's constant, respectively.

In virtue of the first Friedmann equation
\begin{equation}
3 H^{2} \, + 3 \frac{k}{a^{2}} = 8 \pi G \, \rho \, ,
\label{eq:friedmann1}
\end{equation}
where $\rho$ is the  total energy density, we can write
\begin{equation}
{\cal A} = 4 \pi \, \tilde{r}_{A}^{2} = \frac{3}{2 G}\,
\frac{1}{\rho} \, ,\label{harea}
\end{equation}
as well as
\begin{equation}
{\cal A}' = \frac{9}{2G} \, \frac{1\, +  \,w}{a \, \rho} \, ,
\label{primeA1}
\end{equation}
where we have used the conservation equation $\rho' = - 3 (1\, +
\, w) \rho/a$, where $w = p/\rho$ denotes the overall equation of
state parameter, i.e., not just of dark energy. From
(\ref{primeA1}) the area will augment in expanding universes if $1
+ w > 0$ and decrease otherwise.

A further derivation with $w =$ constant yields
\begin{equation}
{\cal A}'' = - \frac{9}{2G}\, \frac{1 \, + \, w}{(a \,
\rho)^{2}}\, (a\, \rho' \, + \, \rho) =  \frac{9}{2G \, a^{2}\,
\rho } \, (1 \, + \, w) \, (2 \, +\, 3w) \, . \label{2primeA1}
\end{equation}
Accordingly, $ {\cal A}'' \leq 0$ for $-1 \leq w \leq -2/3$, and $
{\cal A}''  > 0$ otherwise. Thus, the above criterion disfavor the
dominance of fluids at late times such that the overall equation
of state is either of phantom type or larger than $-2/3$.

We note in passing that when $w \neq$ constant last equation
generalizes to
\begin{equation}
{\cal A}'' =  \frac{9}{2G \, a \, \rho } \, \left[ w' \, + \,
\frac{(1\, + \, w)\, (2\, + \, 3 w)}{a} \right]\, .
\label{2primeAwvar}
\end{equation}

Let us consider the entropy associated to the fluid enclosed by
the apparent horizon. If the fluid is just dust (pressureless
matter, subscript $m$), i.e. $\rho = \rho_{m}$ and $p = p_{m} =
0$, the unphysical result that the temperature vanishes follows
whence Euler's equation, $T \, s = \rho \, + \, p$, cannot be used
to determine the entropy of cold matter (i.e., dust). We proceed
instead as follows. Every dust particle contributes a given bit
-say $k_{B}$- to the fluid entropy. So, within the apparent
horizon we will have $S_{m} = k_{B} \, N \,$ being $\, N = (4
\pi/3)\tilde{r}_{A}^{3} n $  the number of particles there, and $n
= n_{0} \, a^{-3}$ the (conserved) number density of dust
particles. Then,
\begin{equation}
S_{m} =  k_{B}\,  \frac{4 \pi}{3}\, \tilde{r}_{A}^{3} \, n_{0}\,
a^{-3} \propto a^{3/2} \, . \label{eq:Smatter1}
\end{equation}
Hence $S''_{m} \propto \frac{3}{4\, \sqrt{a}} > 0.$  We note
parenthetically that the entropy augments in the volume enclosed
by the horizon because the latter encompasses more and more
particles as the universe expands (i.e., as $H$ diminishes). Thus,
$S''_{m} \, + \, S''_{A} > 0$.

We may now readily understand why sooner or later the Universe
should accelerate, i.e., why it must be endowed with dark energy
(subscript $x$, equation of state $-1 \leq w_{x} \leq -2/3$), or
something dynamically equivalent at the background level (as a
suitably modified gravity). We note parenthetically that, at late
times, dark energy is to dominate over all other energy components
whereby $w \simeq w_{x}$ as $a \rightarrow \infty$. Were the
Universe dominated by radiation and/or matter for ever, $S_{A}''$
could never become negative. And if the horizon entropy dominated
the total entropy, then the Universe  would never tend to a state
of maximum entropy (compatible with the constraints of the
system).

\begin{figure}[htb]
\centering
\includegraphics[width=10cm]{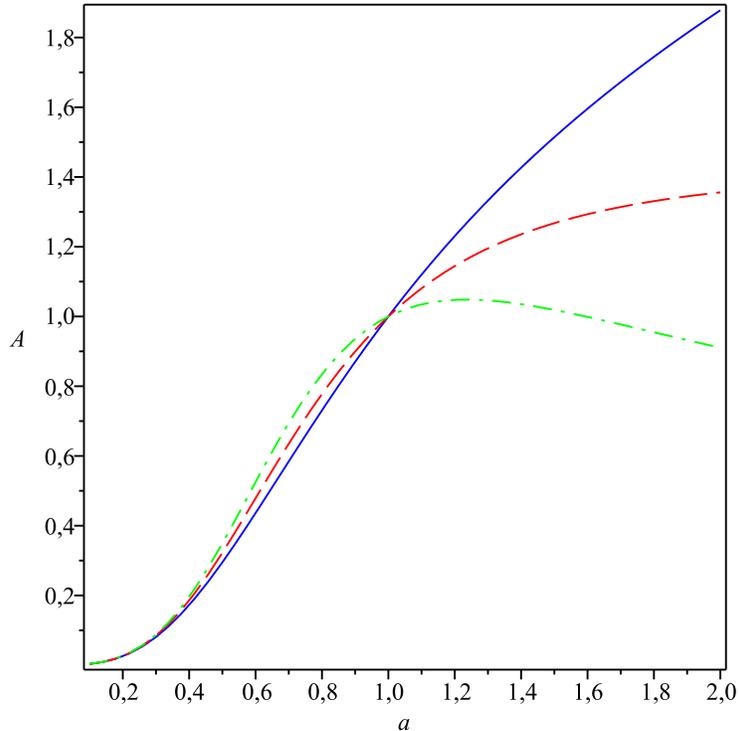}
\caption{Schematic evolution of the area of the apparent horizon
in a FRW universe dominated by cold matter and dark energy with
$w_{x} = -5/6$ (solid line), $w_{x} = -1$ (dashed), and $w_{x}
=-1.2 $ (dot-dashed). In plotting the graphs we have taken
$\Omega_{m0} = 0.3$.} \label{fig:threegraphs}
\end{figure}

Figure \ref{fig:threegraphs}  sketches the evolution of the
horizon area for a spatially flat FRW universe dominated by a
mixture of cold matter and dark energy, i.e., ${\cal A} \propto
[\Omega_{m0} \, a^{-3} \, + \, \Omega_{x0} a^{-3(1+w_{x})}]^{-1}$
for $w_{x}= -5/6$ (solid line), $w_{x} = -1$ (dashed), and $w_{x}
= -1.2$ (dot-dashed) -in the three cases we have assumed
$\Omega_{m0} = 0.3$. It is readily seen that for  $-1 \leq w_{x} <
-2/3$ (solid and dashed lines) the curvature evolves from positive
to negative values and stays negative forever. By contrast, when
the dark energy component is phantom, $w_{x} < -1$ (dot-dashed
line) the curvature goes from positive to positive values with an
intermediate, transient, phase of negative values about the
maximum. Clearly, while in the two first instances the Universe
tends to thermodynamic equilibrium at late times in the case of
phantom it does not. Needless to say, in the absence of dark
energy (not shown) we would have $ {\cal A}'' >0 $ for whatever
value of the scale factor.

\begin{figure}[htb]
\centering
\includegraphics[width=10cm]{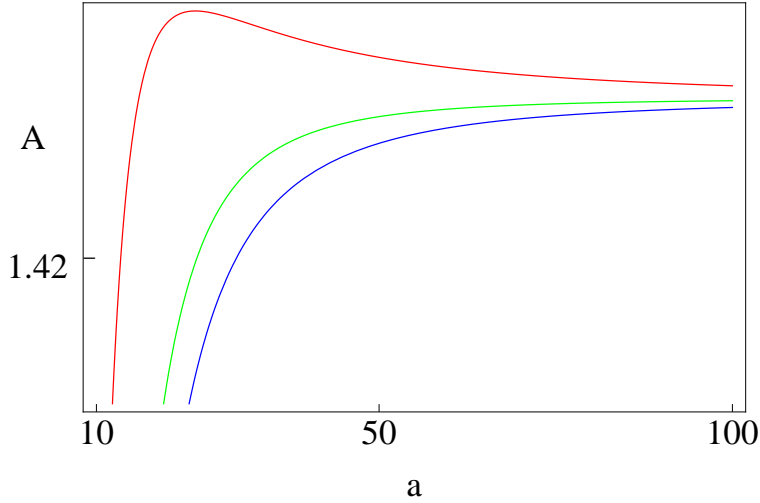}
\caption{Schematic evolution of the area of the apparent horizon
in a FRW universe dominated by cold matter and cosmological
constant ($w_{x} =-1$). From top to bottom, $\Omega_{k0} = -0.02,
\, 0$, and $0.009$, respectively, -see \cite{komatsu}. For $a$
values of the order unity and below (not shown) the graphs
practically overlap each other. In plotting the graphs we have
chosen $\Omega_{\Lambda 0} = 0.7$.} \label{fig:lambda-omegak}
\end{figure}

For completeness, Fig. \ref{fig:lambda-omegak} illustrates the
influence of the spatial curvature on the evolution of ${\cal A}$
when $a \gg 1$ for three $\Lambda$CDM models ($w_{x} = -1$). As
can be seen, the curvature of the graph for positively curved
models remains positive (recall that $\Omega_{k} \equiv - k/(a^{2}
H^{2})$).


We believe that the GSL alongside the criterion that the total
entropy $S$ must fulfill $S'' < 0 $ at late times may serve to
discard some cosmological models and to set limits on the
evolution of $w_{x}$.

\begin{figure}[htb]
\centering
\includegraphics[width=10cm]{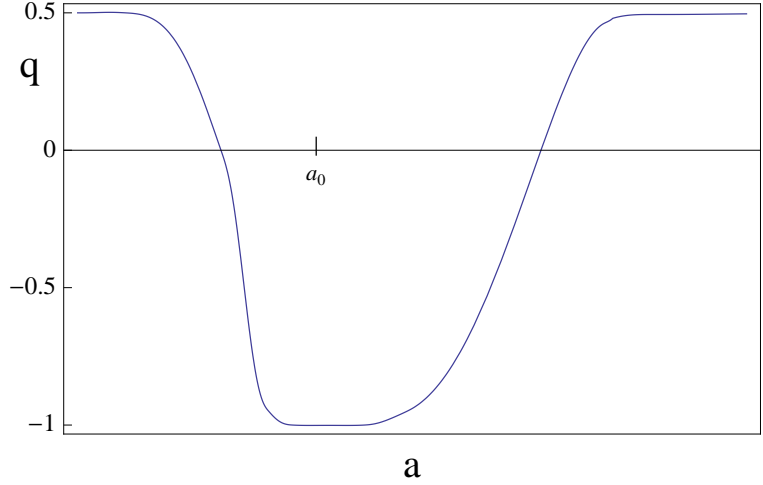}
\caption{Schematic of the evolution of the deceleration parameter,
$q(a)$, given by Eq. (\ref{decp}), in a FRW universe where the
present acceleration stage is transitory and reverts for ever to
matter domination.} \label{fig:q(t)}
\end{figure}

It will prove useful to express $\cal A$ and its derivatives in
terms of the deceleration parameter, $q \equiv -\ddot{a}/(a \,
H^{2})$. The latter can be written as
\begin{equation}
q = -\left(1 \, + \, \frac{a \, H'}{H}\right) \, . \label{decp}
\end{equation}
In spatially flat FRW universes ${\cal A} = 4 \pi H^{-2}$, whence
${\cal A}' = 8 \pi (1+q)/H^{2} $ and
\begin{equation}
 {\cal A}'' = 2 {\cal A}\, \left[ (1\, + \, q) \, (1\, + \, 2q) \, + \, \frac{q'}{a}\right] \, .
\label{2primeA2}
\end{equation}
It follows that for $q \geq -1/2$ the second derivative of the
horizon area, $\, {\cal A}''$, cannot go from positive to negative
values unless $q' < 0$. This excludes evolutions of $q$ as in Fig.
\ref{fig:q(t)}. The latter corresponds to cosmological models in
which the current accelerated expansion is just transitory -after
a period of dark energy dominance, pressureless matter is assumed
to take over \cite{jcap-julio2010,bose-majumdar}.

At this stage it seems fitting to ask ourselves whether the second
derivative of the entropy of the energy components -that for
simplicity we will call fluids, subscript $f$- enclosed within the
apparent horizon will be positive enough so that the sum
$S''_{f}\, + \, S''_{A}\, $ be positive. However, before
proceeding we wish to remark that if the energy component is a
scalar field in a pure quantum state, or the cosmological
constant, it will have no entropy at all. Nevertheless, we wish to
consider the possibility of the scalar field being in a mixture
state and accordingly entitled to have entropy (obviously, this
excludes the case $w_{f} = -1$).

For simplicity we will take $k = 0$ and a single fluid component
with $w_{f} =$ constant. The entropy of the fluid filling the
volume enclosed by the apparent horizon can be estimated by virtue
of Gibbs equation,
\begin{equation}
T_{f} \, dS_{f} = d \left(\rho_{f} \frac{4 \pi}{3}\,  H^{-3}
\right) \, + \, w_{f} \rho_{f} \; d \left(\frac{4 \pi}{3}\,
H^{-3} \right) \, . \label{eq:gibbs}
\end{equation}
With help of (\ref{eq:friedmann1}) and noting that $\rho =
\rho_{f}$
\[
2 G T_{f} \, \frac{dS_{f}}{da} = - \frac{1}{H^{2}} \, (1+3w)
\frac{dH}{da}\, .
\]
Using the second Friedmann equation,
\begin{equation}
 \frac{dH}{da} = - 4 \pi G \, (1\, + \, w_{f}) \, \frac{\rho_{f}}{a
\, H} \, , \label{eq:friedmann2}
\end{equation}
and Eq. (\ref{eq:friedmann1}) once more, we get
\begin{equation}
2 G T_{f} \, \frac{dS_{f}}{da} = \frac{3}{2}\,
(1+w_{f})\,(1+3w_{f}) \frac{1}{aH} \, . \label{1primeSf}
\end{equation}
Thus, $dS_{f}/da$ will be negative or nil for $\, -1 \leq w_{f}
\leq -1/3$ and positive otherwise.

From last equation we can write,
\[
2 G \, \frac{d^{2} S_{f}}{da^{2}} = \frac{3}{2} \, (1+w_{f})\,
(1+3w_{f}) \frac{d}{da}(a\, H \, T_{f})^{-1} \, .
\]
The evolution of the temperature of a perfect fluid, $d \ln
T_{f}/d \ln a = -3 w_{f}$,  readily follows from Gibbs' equation
and the condition for $dS_{f}$ to be a differential expression
-see e.g. \cite{mauricio,grg-db}. Recalling that $w_{f} =$
constant, it leads to $ T_{f} = T_{f0} \, a^{-3w_{f}}$. Using now
$\rho_{f} = \rho_{f0} \, a^{-3(1+w_{f})}$ we arrive to
\begin{equation}
2 G \frac{d^{2}S_{f}}{da^{2}} = \frac{3}{4T_{f0}\, H_{0}} \,
(1+w_{f})\,(1+3w_{f}) \, (1+9w_{f})\, a^{(9w_{f}-1)/2} \,
.\label{d2Sfluid}
\end{equation}
For $-1 < w_{f} < -1/3$ (dark energy) as well as for $w_{f} > -1/9
$ one follows that $S''_{f} > 0$. However, $ S''_{f}$ decreases
with expansion for $w_{f} < 1/9$. The question arises whether the
sum $S''_{f}\, + \, S''_{A}\, $ will be positive or negative in
the long run. To answer it we inspect the ratio between the right
hand sides of Eqs. (\ref{d2Sfluid}) and (\ref{2primeA1}) and note
that $w = w_{f}$,
\begin{equation}
\frac{S''_{f}}{S''_{A}} \propto a^{3(w_{f}-1)/2} \, .
\label{eq:primeSratio}
\end{equation}
This expression vanishes for $w_{f} < 1$  as $a  \rightarrow
\infty$, whereby $S''_{f}\, + \, S''_{A}\, < 0$ in the long run
provided $-1 < w_{f} < -2/3$, i.e., if the fluid is dark energy.

It remains to be seen whether the GSL, $S'_{f} \, + \, S'_{A} \geq
0$, is fulfilled. Note that in virtue of Eqs. (\ref{primeA1}) and
(\ref{1primeSf}), with $T_{f} \propto a^{-3w_{f}}$, the ratio
\begin{equation}
\frac{S'_{f}}{S'_{A}} \propto \frac{1/(aT_{f}\, H)}{1/a \rho_{f}}
\propto
 \rho_{f}^{1/2}\, a^{3w_{f}} \propto a^{3(w_{f}-1)/2}
 \label{eq:2primeSratio}
\end{equation}
vanishes in the long run provided $w_{f} < 1$. Thus albeit
$S'_{f}$ is negative for $-1 < w_{f} <-1/3$, the GSL is satisfied
for $w_{f} > -1$ when $\, a\rightarrow \infty$.

Altogether, dark energy with constant equation of state in the
interval $(-1, -2/3)$ respects the GSL as well as the criterion
that $S''_{f} \, + \, S''_{A} < 0$ when $a \rightarrow \infty$.
This result is consistent with the tightest observational
constraints available -cfr. Table in IV in \cite{komatsu}. In our
view, cosmological models that meet both criteria should be
preferred to those failing any of the two.

\subsection{Parameterized $w_{x}$ models}
In this subsection we consider the model of Barboza and Alcaniz in
which the dark energy equation of state is parameterized in terms
of  redshift as \cite{Barboza-2008}
\begin{equation}
w_{x}(z) = w_{0}\, + \, w_{1} \frac{z(1\, + \, z)}{1\, + \, z^2}
\, . \label{eq:barboza}
\end{equation}
It has the advantage over other parameterizations of not diverging
at any time. We recall, parenthetically,  that the redshift is
related to the scale factor by $ 1\, + \, z = 1/a \, $.

The pair of free, constants parameters $w_{0}$ and $w_{1}$ in
(\ref{eq:barboza}) ought to be restricted by physical requirements
and observation. A first constraint $ w_{0} \, +w_{1} < 0 $
follows from demanding that dark energy be subdominant at early
times (when $z \gg 1$), otherwise cosmic structure would never had
formed. Next, we will use the GSL together with the condition that
the sum $\, S''_{m} \, +\, S''_{x} \, + \, S''_{A} \,$ be negative
or nil to set further constraints on the said pair.

We begin by writing the Hubble function of a spatially-flat FRW
universe dominated by cold matter and dark energy, with equation
of state (\ref{eq:barboza}),
\begin{equation}
H^2=H_0^2\left[\Omega_{m0}a^{-3}+\Omega_{x0}
a^{-3(1+w_0+w_1)}(2a^2-2a+1)^{3w_1/2}\right] \, .
\label{barboza-hubble}
\end{equation}

We next consider the entropy of the apparent horizon, $S_{A}
\propto {\cal A}$. Two separate cases arise: (i) $w_1 < 2/3$ and
$-2/3 < w_0 < -w_1$ as well as (ii) $w_0<-2/3$ and $w_0+w_1<0$. In
the first case, $S'_{A}$ grows without bound and we do not
consider it any further. In the second one $S'_{A}$ approaches
zero asymptotically from above. As $\, a \rightarrow \infty$, we
see that $w_{x}(a) \rightarrow w_{0}$, $\Omega_{x} \rightarrow 1$,
$a w'_{x} \rightarrow 0 \; \; (\sim a^{-1})$ and $ H^{2} \sim
a^{-3(w_0+1)}$ whence all the terms but the last one, that decays
as $a^{-1}$, in the second derivative of the horizon entropy,
\begin{equation}\label{S''}
S''_{A} \propto {\cal A}'' \propto\frac{1}{a^2 H^2}\,
\left[3(w_{x} \, \Omega_x)^2+\frac{5}{2} w _{x} \,
\Omega_x-\frac{3}{2}w_{x}^2 \, \Omega_x \, + \, 1+ \, \frac{1}{2}a
\, \Omega_x \, w_{x}'\right] \, ,
\end{equation}
present a like behavior. In consequence, the quantity in the
square brackets tends to a constant, and $S''_{A} \sim
a^{3w_{0}+1}$ as $ a \rightarrow \infty$. For $S''_{A}$ to be
negative that constant must be negative as well, then $ 3 w_{0}^2
\, + \, 5 w_{0}\, + \, 2 < 0 $.

\begin{figure}[htb]
\centering
\includegraphics[width=8cm]{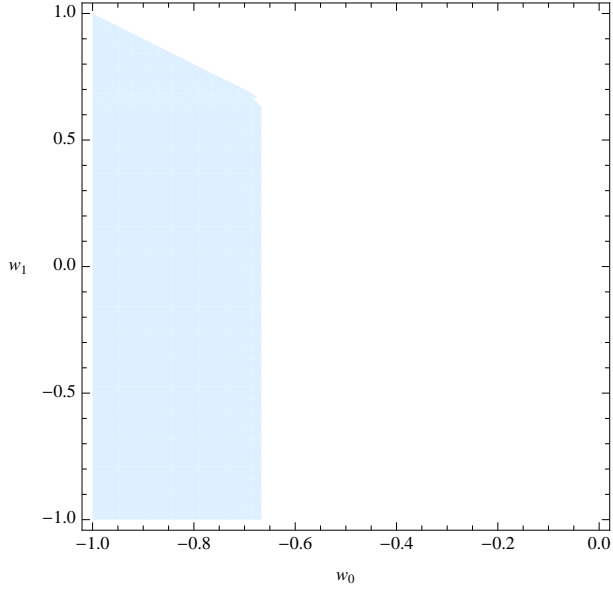}
\caption{The available region in the $(w_0,w_1)$ plane for the
parameterized-$w_{x}$ model of Barboza and Alcaniz
\cite{Barboza-2008}.} \label{fig:planew0w1}
\end{figure}

\noindent Finally, this constraint and the previous one set the
accessible range to
\begin{eqnarray}\label{hor1}
\left\{ \begin{array}{l}w_1<2/3\\-1<w_0<-2/3
\end{array}\right.& \cup&
\left\{ \begin{array}{l}2/3\leq w_1<1\\-1<w_0<-w_1,
\end{array}\right.
\end{eqnarray}
as shown in Fig. (\ref{fig:planew0w1}).

Obviously, when the entropy of matter and dark energy are taken
into account the previous result may vary. With the help of the
equation for the temperature of dark energy
\begin{equation}
T_x(a)=T_{x0} a^{-3(w_0+w_1)}(2a^2-2a+1)^{3 w_1/2} \, ,
\label{t-de-barboza}
\end{equation}
it can be checked that
\[
\frac{S'_{x}}{S'_{m}} \sim C_{1} \qquad
\text{and}\qquad\frac{S''_x}{S''_m}\sim C_{2},
\]
where $C_{1}$ and $C_{2}$ are constants. Hence we focus on matter
contribution with respect to the horizon.

When $w_{0} > 0$ it follows that
\[
\frac{S'_m}{S'_A}\sim \frac{S''_m}{S''_A}\sim a^{-3/2},
\]
whereby the horizon contribution prevails in the long run.
However, we have already discarded this solution because $S_{A}'$
diverges. In a sense, this is a consistency check on our criteria
because observationally $w_{0}$ is known to be negative.

In the opposite case, $w_{0} < 0$, one obtains
\[
\frac{S'_m}{S'_A}\sim \frac{S''_m}{S''_A}\sim a^{3(w_{0}-1)/2}.
\]
Again, the horizon contribution prevails, and the result
(\ref{hor1}) stands.

The FRW universe model dominated by cold matter plus dark energy,
with equation of state parameterized according to Eq.
(\ref{eq:barboza}), has been observationally constrained in
\cite{Barboza-2008} by using data from supernovae type Ia, baryon
acoustic oscillations, and cosmic background radiation. The best
fit values of the parameters at $1\sigma$ confidence level were
found to lie in the ranges
\begin{equation}
-1.35\leq w_0\leq -0.86 \, , \qquad \qquad -0.33\leq w_1\leq 0.91
\, , \label{eq:w0w1range}
\end{equation}
in full consistency with the above result, Eq. (\ref{hor1}).

\subsection{Chaplygin gas model}
The original Chaplygin gas model unifies matter and dark energy in
the sense that they are no longer two separate components but a
unique entity that mimics cold matter at early times and a
cosmological constant at late times \cite{kamenshchik-2001}. Its
equation of state $\, p = -A/\rho$, with $A$ a positive constant
and $\rho$ the energy density of the gas (i.e., the total energy
density), is obtainable from the Nambu-Goto action for a d-brane
moving in the $d+1$ dimensional bulk \cite{Jackiw-2000}. In a FRW
universe, the dependence of the energy density  on the scale
factor reads $\, \rho = \sqrt{A \, + \, (B/a^{6})}$, where $B$ is
nonnegative integration constant; or equivalently,
\begin{equation}
\rho = \rho_{0} \left[ 1\, - \, \Omega_{\star} \, + \,
\Omega_{\star} \, a^{-6} \right]^{1/2}
\label{eq:hubble-chaplygin}
\end{equation}
with $\Omega_{\star} = B/(A+B)$.

The evolution of the horizon area (${\cal A} \propto 1/\rho$) is
akin to the one depicted by the solid line in Fig.
\ref{fig:threegraphs}; that is to say, it grows monotonously and
presents negative curvature at sufficiently large scale factor.

By contrast, since the entropy of the Chaplygin gas obeys $S_{Ch}
\propto (1\, + \, w)/(HT)$, where
\begin{equation}
w = \frac{p}{\rho} = - \left[1 \, + \, \frac{\Omega_{\star}}{1 \,
- \, \Omega_{\star}}\, a^{-6} \right]^{-1} \, ,
\label{eq:w-chaplygin}
\end{equation}
refers to the equation of state parameter of the gas (which
evolves from zero to $-1$), and the temperature is governed by
$T'/T = -3 w/a$, one follows that $S'_{Ch} < 0$ and $S''_{Ch} > 0$
for large scale factor. However, as it can be checked,
$S'_{Ch}/S'_{A} \rightarrow 0 $ and $S''_{Ch}/S''_{A} \rightarrow
0 $ as $a \rightarrow \infty$. Accordingly, the total entropy, $S
= S_{Ch} \, + \, S_{A}$, fulfills $S' > 0$ and $S''< 0$ in the
same limit. In other words, the GSL is respected in that limit and
the Universe tends to thermodynamic equilibrium in the long run.

\subsection{Holographic, interacting models}
Here we first consider a spatially-flat, holographic, interacting
model dominated by pressureless matter and dark energy. The latter
component is assumed holographic in the sense that its energy
density varies as the area of the apparent horizon and interacts
with matter at a given, non-constant, rate. As a result $w_{x}$
decreases with expansion and the fractional densities of both
components remain constant. This much alleviates the cosmic
coincidence problem \cite{cqg-wd} and the model shows
compatibility with observation \cite{jcap-ivan}.

Inspection of the Hubble factor, given by Eq. (2.4) of
\cite{jcap-ivan},
\begin{equation}
H(a) = H_{0} \, \left[\frac{\Gamma}{3H_{0}\, r}\, + \, \left( 1\,
-\, \frac{\Gamma}{3H_{0}\, r}\right)\, a^{-3/2}\right] \, ,
 \label{eq:H(a)ivan}
 \end{equation}
readily reveals that the graph of the evolution of the area of the
apparent horizon, ${\cal A} \propto 1/H^{2}$, has positive
curvature at the beginning of the expansion and  gently evolves to
negative curvature values remaining thus for ever, similarly to
the solid line in Fig. \ref{fig:threegraphs}, for the best fit
observational values, $\Gamma/H_{0} = 0.563$ and $r = 0.452$,
obtained in \cite{jcap-ivan}.

As can be readily checked, the second derivative of the entropy of
the dark energy  within the horizon, $S_{x}$, is negative as well.
Effectively,
\begin{equation}
S_{x} = \frac{4 \pi}{3} \tilde{r}_{A}^{3} \, s = \frac{4 \pi}{3}
\, \frac{1}{H^{3}} \, \frac{1+w_{x}}{T_{x}} \, \rho_{x} \propto
\frac{1+w_{x}}{H \, T_{x}} \, , \label{eq:Sx-holographic}
\end{equation}
where $r = \rho_{m}/\rho_{x}$ and $T_{x} \propto a^{-3w_{x}}$ when
$a \rightarrow \infty$. For $a \gg 1$ one has that $w = {\rm
constant}$, and one obtains $S''_{x} \propto 3(1+w_{x})\, w_{x} \,
(3w_{x}-1) < 0$ since, in the model at hand, $1+w_{x} < 0$.\\
As for the entropy of pressureless matter
\begin{equation}
S_{m} =  k_{B}\,  \frac{4 \pi}{3}\, \tilde{r}_{A}^{3} \, n =
k_{B}\,  \frac{4 \pi}{3}\, \frac{\rho_{m}}{m \, H^{3}} = k_{B}\,
\frac{4 \pi}{3} \frac{1}{m \, H^{3}}\, \frac{r}{1+r} \, (\rho_{m}
+ \rho_{x}) \propto \frac{1}{H} \, , \label{eq:Smatter2}
\end{equation}
where $m$ stands for the mass of the particles. (Recall that in
this model $r = {\rm constant}$). We then have, $S'_{m}  \propto
-H'/H^{2}$ and $S''_{m} \propto 2 (H'^{2}/H^{3})\, -\,
(H''/H^{2})$. Using (\ref{eq:H(a)ivan}) we find that $S'_{m} > 0$
and
\[
S''_{m} \propto \gamma \, \beta^{2} \; \frac{\gamma -5 \,
a^{3/2}}{a^{5}\, H^{3}} \, ,
\]
where $\beta$ is a short-hand for the dimensionless combination
$\Gamma/(3H_{0} r)$ and $\gamma = (1-\beta)/\beta$; both
quantities lie in the range (0,1). For $a \gg 1$, the second term
is the leading one whence, at sufficiently large scale factor,
$S''_{m}$ is also negative. In summary, in this model, $ S''_{A}\,
+ \, S''_{m} \, + \, S''_{x}  < 0$ for $a \rightarrow \infty$.

\begin{figure}[htb]
\centering
\includegraphics[width=10cm]{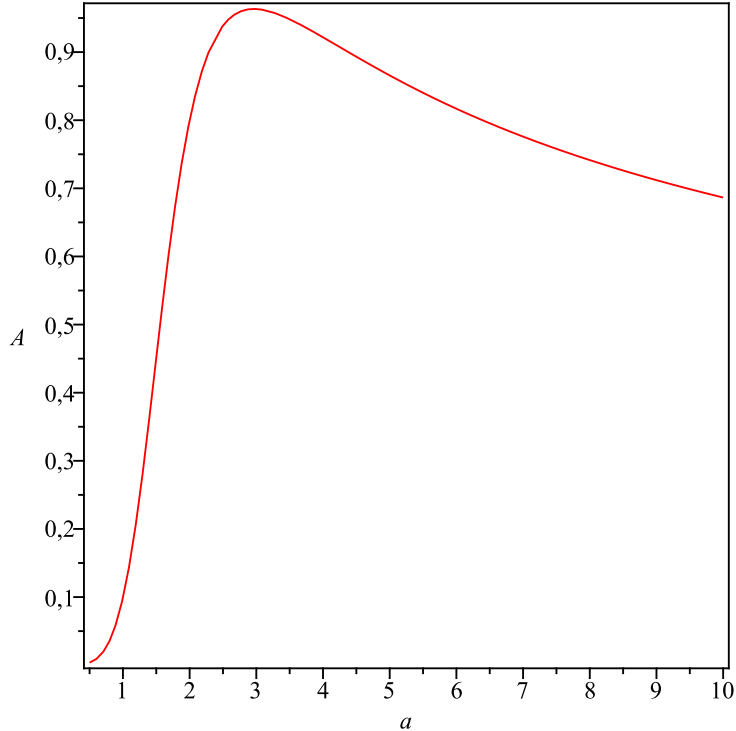}
\caption{Schematic evolution of the area of the apparent horizon
of the holographic, non-interacting model of Gao {\it et al.}
\cite{prd-gao}. The curvature is negative about the maximum and
becomes larger than zero later on and remains thus for ever. In
drawing the graph we have adopted the values in the said
reference: $\Omega_{k0} = 0 $, $\Omega_{r0} = 8.1 \times 10^{-5}$,
$\Omega_{m0} = 0.27$, $\Omega_{x0} = 0.73$, $\alpha = 0.46$, and
$f_{0} = 0.65$.} \label{fig:gao-holographic}
\end{figure}

It would be misleading, however, to believe that for all
holographic models  ${\cal A}''< 0$ when $a \gg 1$. For instance,
in the holographic model of Gao {\it et al.} \cite{prd-gao}, which
uses the Ricci's length as infrared cutoff, the Hubble equation
reads
\begin{equation}
H = H_{0} \, \sqrt{\Omega_{k0} \, a^{-2} \, + \, \Omega_{m0} \,
a^{-3} \, + \, \Omega_{r0} \, a^{-4} \, + \,
\frac{\alpha}{2-\alpha} \, \Omega_{m0} \, a^{-3} \, + \, f_{0} \,
a^{(2/\alpha)-4}} \, , \label{eq:hubble-gao}
\end{equation}
where the subscript $r$ stands for radiation; $\alpha \simeq 0.46$
and $f_{0}\simeq 0.65$ are dimensionless parameters. As Fig.
\ref{fig:gao-holographic} shows, the area of the apparent horizon
grows to a maximum to monotonously decrease afterwards for ever.
Consequently, ${\cal A}''> 0$ as $a \rightarrow \infty$.

On the other hand, the entropies of the fluid components
(radiation, matter, and dark energy) decrease with expansion and
their second derivatives are positive for large scale factor.
Thus, this model violates the GSL and does not approach
thermodynamic equilibrium at late times.

\section{Modified gravity models}
Models that depart from Einstein gravity may lead to a late
acceleration era without the help of any exotic component of
negative pressure. Here we examine some of them, namely: the one
based in the brane-induced gravity model of Dvali {\it et al.}
\cite{dvali00} (see \cite{deffayet01,deffayet02a}), the original
Cardassian model proposed by Freese and Lewis \cite{cardassian},
and the torsion model of Bengochea and Ferraro
\cite{bengochea-2009}, to check whether they fulfill these
criteria. Before we proceed, it is fair to say that, as far as we
know,  it has not been rigorously proven that the entropy
associated to the  apparent horizon of the two latter models
simply obeys $S_{A} \propto {\cal A}$, but it seems to us a very
reasonable assumption and we shall adopt it.

\subsection{Dvali-Gabadazze-Porrati's model}
%
This model considers our 4-dimensional Universe as a brane
embedded in a 5-dimensional bulk with flat Minkowski metric. As a
consequence of the brane-induced term, the conventional
Friedmann's equation modifies to
\begin{equation}
H^{2} \, + \, \frac{k}{a^2} = \left(\sqrt{\frac{\rho}{3M_{Pl}^2}\,
+ \, \frac{1}{4r^2_c}}\, + \, \frac{1}{2r_{c}}\right)^2 \, ,
\label{eq:hubble-dgp}
\end{equation}
where $\rho$ and $r_{c}$ stand for the total energy density
(matter plus radiation in this model) and the crossover scale
below which gravity appears as four dimensional, respectively.

\begin{figure}[htb]
\hspace{-4 cm}
\begin{minipage}{0.3\textwidth}
\centering
 \includegraphics[width=8cm]{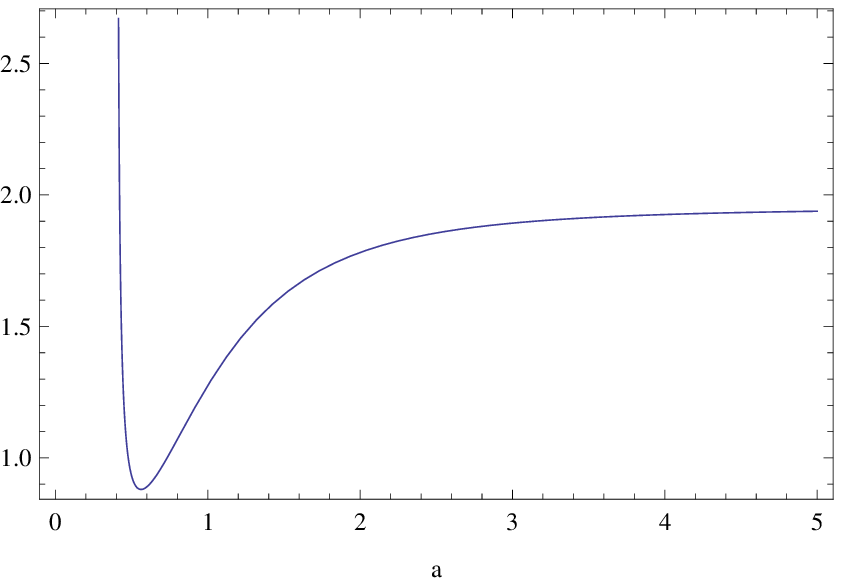}
 \end{minipage}
\hspace{3 cm}
\begin{minipage}{0.3\textwidth}
\centering
 \includegraphics[width=8cm]{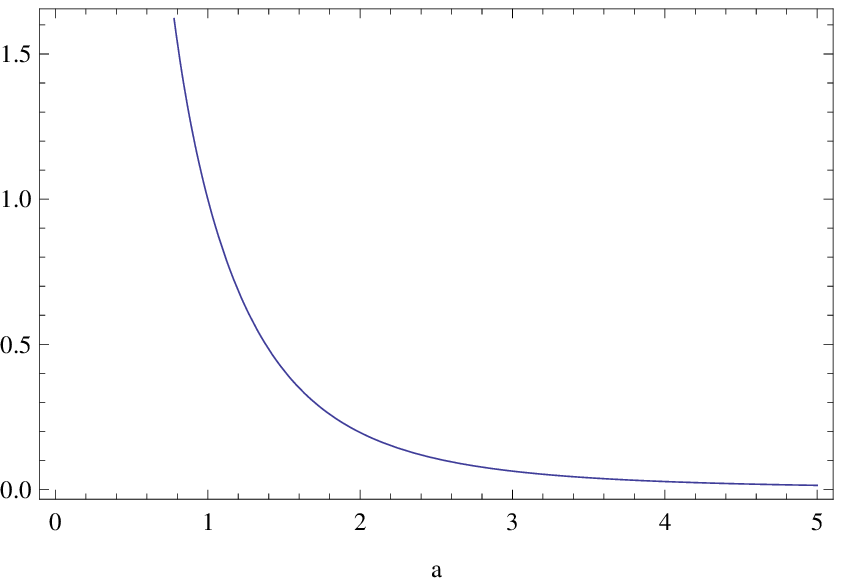}
  \end{minipage}
  \caption{Left panel: Schematic evolution of the entropy of the horizon of the
  Dvali-Gabadazze-Porrati model. Right panel: The single graph depicts the qualitative
  evolution of the entropy of radiation as well as of cold matter inside the horizon.
  In drawing the graphs we have used the best-fit values of the parameters of the
  model, $r_{c}\, H_{0} = 1.21$ and $\Omega_{m0} = 0.18$, obtained in
  \cite{deffayet02b}.} \label{fig:DGP}
\end{figure}

The entropy of the apparent horizon, as computed by Sheykhi {\it
et al.} \cite{sheykhi07},
\begin{equation}
S_{A} = k_{B}\, \frac{3\pi \, \tilde{r}^{2}_{A}}{\ell^{2}_{pl}}\,
\left[1 \, + \, \frac{\tilde{r}_{A}}{r_{c}}\right] \, ,
\label{eq:SA-DGP}
\end{equation}
is no longer proportional to the horizon area but contributed by
two terms. The first one corresponds to the Bekenstein-Hawking
entropy and accounts for gravity on the brane. The second one is
related to the bulk. Its evolution in terms of the scale factor is
sketched in the left panel of Fig. \ref{fig:DGP}. From a certain
point on it increases monotonously with negative curvature.
As before, the entropy of dust and radiation are proportional to
$(aH)^{-3}$. Both of them decrease monotonously with positive
curvature. It is found that
\begin{equation}
\frac{S_{m}^{\prime }}{S_{A}^{\prime }}\sim \frac{S_{r}^{\prime }}{%
S_{A}^{\prime }} \, \rightarrow \,  -\frac{1}{%
3\Omega _{m0}r_{c}H_{0}}\,  \label{eq:DGPlimiting1}
\end{equation}
(subscript $r$ for radiation), bear in mind that $r_{c}$ has
physical dimensions of time. Thus, in the long run it leads to a
negative constant -though the latter does not exactly coincides in
the radiation and matter cases.

Once again the question arises whether the total entropy, $S =
S_{m} \, + \, S_{r} \, + \, S_{A}$, obeys  the GSL and presents
negative curvature when $a \gg 1$. To answer this consider the
limiting expressions when $a \rightarrow \infty$,
\begin{eqnarray}
S^{\prime } &=&S_{A}^{\prime }\left[ 1\, + \, \frac{S_{m}^{\prime }}{%
S_{A}^{\prime }}\, + \, \frac{S_{r}^{\prime }}{S_{A}^{\prime }}\right] \rightarrow \ \ 0 \, ,
\label{eq:DGPlimiting2a}\\
S^{\prime \prime } &=&S_{A}^{\prime \prime }\left[ 1 \, + \,
\frac{S_{m}^{\prime
\prime }}{S_{A}^{\prime \prime }}\, + \, \frac{S_{r}^{\prime \prime }}{%
S_{A}^{\prime \prime }}\right] \rightarrow \ \ 0 \, .
\label{eq:DGPlimiting2b}
\end{eqnarray}
The first one tends to zero from above, the second one from below
(following the healthy behavior of the apparent horizon entropy).
Both relations when taken together imply the inequality
\begin{equation}
1-\frac{4 \ell_{Pl}^{2}c}{27k_{B}\Omega _{m0}r_{c}H_{0}^{2}}\left[ k_{B}n_{0}+4%
\frac{\rho _{r0}c^{2}}{T_{r0}}\right] \, ,
\label{eq:DGPinequality1}
\end{equation}
which translates into an upper bound on the current number density
of dust particles,
\begin{equation}
n_{0}<\frac{1}{k_{B}}\left[ \frac{27k_{B}\Omega _{m0}r_{c}H_{0}^{2}}{%
4 \, \ell_{Pl}^{2}c} \right] \sim 10^{38} \;  {\rm cm}^{-3} \, .
\label{eq:DGPinequality2}
\end{equation}
Since it is fulfilled by a huge margin the GSL is satisfied and
$S''$ results negative in the long run.

\subsection{Cardassian model}
In this spatially-flat FRW model the first Friedmann equation
acquires an extra term that accounts for acceleration at
sufficiently high redshifts,
\begin{equation}
H^{2} = \frac{8 \pi G}{3} \, \rho \, + \, B \rho^{\alpha} \, .
\label{eq:cardassian}
\end{equation}
Here two new non-negative constants, $B$ and $\alpha$, appear
while $\rho$ stands for the energy density of cold matter -the
only energy component. For $\alpha < 2/3$ the universe features a
transition from $q > 0$ to $q < 0$ at redshift
\[
z_{tr} = \left[\frac{(1-\frac{3 \alpha}{2})\, B}{(4 \pi G/3)\,
\rho_{0}^{1-\alpha}} \right]^{1/3(1-\alpha)} \, ,
\]
without need whatsoever of dark energy.

As can be checked, ${\cal A} = 1/H^{2}$ evolves in such a way that
while ${\cal A}' > 0 $ at all redshifts,  ${\cal A}''$ changes
from positive to negative values at some point and stays thus for
ever. This was to be expected since the extra term on the right
had side of (\ref{eq:cardassian}) dynamically amounts to the
presence of some dark energy component. Thus, the overall behavior
of ${\cal A}$ should be qualitatively similar to that of a model
dominated by a mixture of pressureless matter and dark energy. To
be more specific, comparison of the right hand sides of
(\ref{eq:cardassian}) and the first Friedmann equation for a
mixture of cold dark matter and dark energy with constant $w_{x}$
and $k = 0$, namely, $H^{2} = (8 \pi G/3)[\rho_{m0} \, a^{-3} \, +
\, \rho_{x0} \, a^{-3(1+w_{x})}]$, shows that, at the background
level, every Cardassian model can be mapped to a spatially-flat
dark energy one satisfying $\alpha = 1\, +\, w_{x}$ and $B = (8
\pi G/3)\,(\rho_{x0}/\rho_{m0}^{\alpha})$. As a consequence ${\cal
A}'' < 0$ for sensible $ \alpha$ values (i.e., $0 < \alpha <
2/3$).

Although the entropy of the matter enclosed by the apparent
horizon varies as $(a\, H)^{-3}$, whence $S'_{m} < 0$ and $S''_{m}
> 0$ for $a \gg 1$, the ratios $S'_{m}/{\cal A}' \propto
a^{3(w_{x}-1)/2}$ and $S''_{m}/{\cal A}'' \propto
a^{3(w_{x}-1)/2}$ tend to zero in the same limit. It implies that
$S'_{m} \, + \, S'_{A} > 0 $ and $S_{m}'' \,  + \, S_{A}'' < 0$ as
$a \rightarrow \infty$.

\subsection{Torsion model}
The torsion model of Bengochea and Ferraro \cite{bengochea-2009}
falls into a class of models that describes gravitation in terms
of the torsion scalar $\tau$ rather than curvature, $R$. Thus, the
action takes the form
\begin{equation}
I = \frac{1}{16\pi G}\int d^4 x \sqrt{-g} \left(\tau +
f(\tau)\right)+ I_{matter} \, , \label{eq:Itorsion}
\end{equation}
where $f(\tau)$ is a free function to be constrained by
observation and experiments. An important advantage of this set of
theories is that the field equations are second  order as opposed
to the four order equations of $f(R)$ gravity. However, up to now,
specific models have been proposed in the cosmological context
only. Works investigating the symmetries and dynamics of the
theory show that it can exhibit extra degrees of freedom since the
theory is not local Lorentz invariant \cite{jdbarrow-2010}.

Bearing in mind that for the FRW metric $\tau = -6H^{2}$, from
(\ref{eq:Itorsion}) the generalized Friedmann equations
\begin{eqnarray}
H^2&=& \frac{8\pi G}{3}\rho -\frac{f}{6}-2 H^2 f_{\tau} \, , \label{eq:torsion-friedmann1} \\
\frac{d H^2}{d\ln{a}}&=&\frac{16 \pi G p + 6H^2+f+12 H^2
f_{\tau}}{24 H^2 f_{\tau \tau}-2-2 f_{\tau}} \, ,
\label{eq:torsion-friedmann2}
\end{eqnarray}
follow. Here, $\rho$ and $p$ denote the energy density and
pressure of the fluid component -which, we will assume, just cold
matter.

In the model of Ref. \cite{bengochea-2009}
\begin{equation}
f(\tau) = -\alpha(-\tau)^{-n} \, , \label{ftau}
\end{equation}
where $\alpha=(1-\Omega_{m0})(6H_{0}^{2})^{1-n}/(2n-1)$.

\begin{figure}[htb]
  \hspace{-4 cm}
\begin{minipage}{0.3\textwidth}
\centering
 \includegraphics[width=8cm]{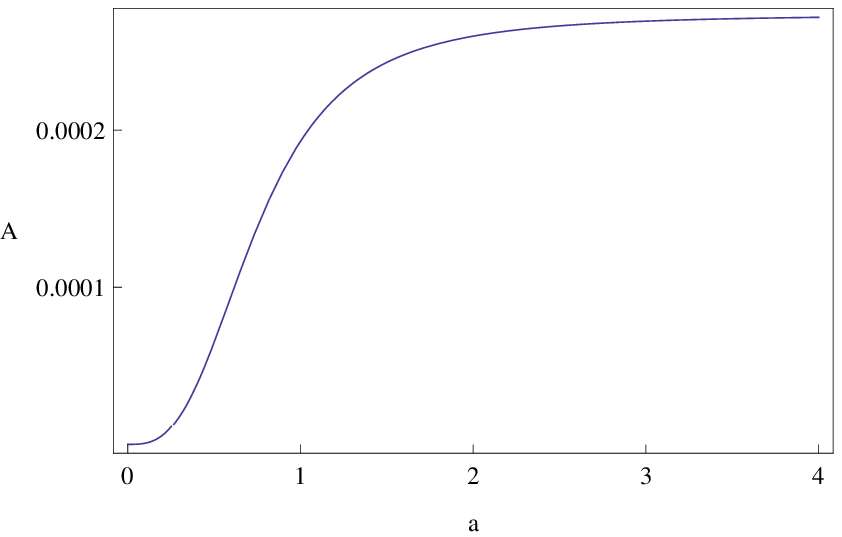}
 \end{minipage}
 \hspace{3 cm}
\begin{minipage}{0.3\textwidth}
\centering
 \includegraphics[width=8cm]{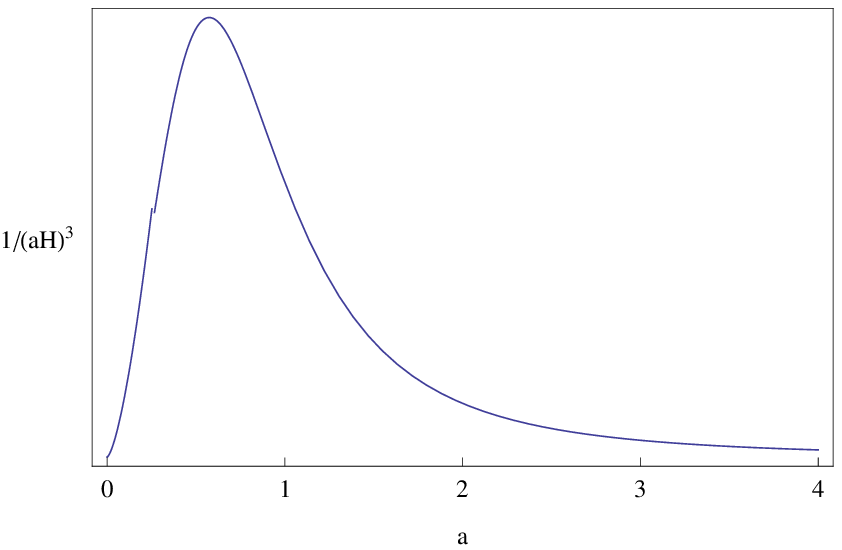}
  \end{minipage}
 \caption{Schematic evolutions of the area of the
 apparent horizon in the case of the $f(\tau)$ model of
 Ref. \cite{bengochea-2009} (left panel), and the entropy of
 the cold matter enclosed by the apparent horizon (right panel). In plotting the
graphs we employed the best fit values of the said model, $n=-0.1$
and $\Omega_{m0}=0.27$.}
\label{fig:Af(T)}
\end{figure}

We evaluate the entropy of the apparent horizon and matter entropy
using the best fit values of the parameters, $n = -0.1$ and
$\Omega_{m0} = 0.27$ -cfr. Ref. \cite{bengochea-2009}. Left panel
of Fig. \ref{fig:Af(T)} shows the evolution of area of the
apparent horizon in terms of the scale factor. The plot starts
with positive curvature and is ever increasing. Later on, the
effect of the ``dark torsion" is felt and the sign of the
curvature changes to remain negative for ever. The right panel
qualitatively depicts the evolution of the entropy of cold matter
(proportional to $(aH)^{-3}$). It begins increasing to decrease
after the maximum. The curvature is negative about the maximum
only. Thus, the question arises whether for $a \rightarrow \infty$
the GSL will be respected and the sum $ S''_{m} \, + \, S''_{A}$
will be negative. A numerical study shows that the ratios
$|S'_{m}|/S'_{A}$ and $S''_{m}/|S''_{A}|$ increase with the scale
factor when $a \gg 1$. Therefore the answer to both questions
appears to be no.

Before closing this subsection it is fair to recall that it is
still unknown whether $f(\tau)$ models admit black holes
solutions. If they don't, it will become unclear that some
connection entropy-horizon area really holds in these theories.

\section{Discussion}
As we have argued, neither a radiation nor a cold matter dominated
universe can tend to thermodynamic equilibrium in the long run. By
contrast, dark energy dominated universes may; this holds true
irrespective of whether the dark energy component has entropy or
not. Accordingly, dark energy (or something dynamically equivalent
at the background level, such as a suitably modified gravity
theory) appears  thermodynamically motivated. In other words, any
of these two kind of ingredients was to be expected on
thermodynamic grounds. We, therefore, should not wonder that the
Universe is accelerating. However, it does not mean that every
accelerating universe is thermodynamically motivated; that is the
case, for instance, of any phantom dominated expansion with $w_{x}
= {\rm constant}$, and  some modified gravity models.

One may object that if our reasoning were valid, the Universe
would have never ceased to inflate as it would mean a transition
from acceleration, ${\cal A}'' < 0$, to deceleration ${\cal A}'' >
0$. Clearly if the primordial inflation would have lasted for
ever, big bang nucleo-synthesis would never have occurred,
galaxies couldn't have come into existence, and so on -something
in stark contrast with observation. However, this reasoning is
rather incomplete as it leaves aside the huge entropy generated
during the reheating process at the end of inflation. In this
explosive and quasi-instantaneous event the inflaton field
relinquishes all its energy in the form of matter and radiation
and, as a consequence, the Universe sees its temperature
enormously increased \cite{lyth-liddle}. Since this huge amount of
matter and radiation thermalizes (a necessary condition for
primordial nucleosynthesis) both second derivatives, $S''_{m}$ and
$S''_{r}$, are negative and, as a result, we may well have that
$S''_{r} + S''_{m} + S''_{A} < 0$. Obviously, this will depend on
the specific inflationary model and the particular reheating
process involved, but we are not aware of any  general argument
against the fulfillment of this inequality. Specific calculations
in this connection will be the subject of a future research.

Interestingly enough, cosmological models complying both with the
GSL and the thermodynamic criterion that, at late times, the
Universe should approach thermodynamic equilibrium appear to be
compatible with observational data. Nevertheless, some models that
do not comply with the said criteria (as some phantom models and
some modified gravity models) seem also consistent with the said
data.

Our argument could be falsified if it were discovered that in the
past (but after the radiation-dominated period set in) the
Universe experienced one (or more) transitions from deceleration
to acceleration and back. Cosmologies of the type have been
proposed to explain the seemingly periodic distribution of
galaxies with redshift -see, e.g. \cite{apjl-morikawa},
\cite{grg-nds}- and as an expedient to solve the cosmic
coincidence problem \cite{ruth}. However, recent studies on the
impact of hypothetical transient periods of
acceleration-deceleration on the matter growth \cite{prd_linder}
and on the radiation power spectrum \cite{jcap_linder} from the
decoupling era, $a \simeq 10^{-5}$, to $a = 0.5$ practically
discard these periods in the redshift intervals considered.

If eventually it gets confirmed that the present phase of
acceleration is to last forever, it may be seen as an indication
that the Universe as a whole obeys the laws of thermodynamics
(with the reservation that not all models that accelerate at late
times comply with them). If, on the contrary, the Universe resumed
a decelerated stage one should either call into question the
validity of applying the said laws in a cosmological setting or
wait for a later, and definitive, accelerating era.

Altogether,  it is for the reader to decide which possibility
looks the less queer: dark energy (or modified gravity) or a
universe that will never approach thermodynamic equilibrium. The
remaining possibility, either dark energy or modified gravity
combined with an increasing departure from equilibrium, involves
two oddities rather than one.

\acknowledgments{We are grateful to Fernando Atrio, Bin Wang,
 Emilio Elizalde, David Jou, Jos\'{e} Gaite, Jack Ng, and Winfried Zimdahl
 for comments on the subject of this paper. NR was funded by the
Spanish Ministry of Science and Innovation through the
``Subprograma Estancias de J\'{o}venes Doctores Extranjeros,
Modalidad B", Ref: SB2009-0056. This research was partly supported
by the Spanish Ministry of Science and Innovation under Grant
FIS2009-13370-C02-01, and the ``Direcci\'{o} de Recerca de la
Generalitat" under Grant 2009SGR-00164.}

\end{document}